\documentclass[11pt,a4paper]{article}
\usepackage{jcappub}

\title{Non-perturbative  and self-consistent models of neutron stars in $R$-squared gravity}
\author[a,b]{Stoytcho S. Yazadjiev,}
\author[a]{Daniela D. Doneva,}
\author[a]{Kostas~D.~Kokkotas,}
\author[b]{Kalin V. Staykov\,}

\affiliation[a]{Theoretical Astrophysics, Eberhard-Karls University
of T\"ubingen, T\"ubingen 72076, Germany} \affiliation[b]{Department
of Theoretical Physics, Faculty of Physics, Sofia University, Sofia
1164, Bulgaria}

\emailAdd{yazad@phys.uni-sofia.bg}
\emailAdd{daniela.doneva@uni-tuebingen.de}
\emailAdd{kostas.kokkotas@uni-tuebingen.de}
\emailAdd{kalin.v.staikov@gmail.com}

\abstract{In the present  paper we  investigate non-perturbatively
and self-consistently the structure of neutron stars in $R$-squared
gravity by simultaneously solving the interior and exterior problem.
The mass-radius relations are obtained for several equations of
state and for  wide range of the $R$-squared gravity parameter $a$.
Even though the deviation from general relativity for nonzero values
of $a$ can be large, they are still comparable with the variations
due to different modern realistic equations of state. That is why
the current observations of the neutron star masses and radii alone
can not put constraints on the value of the parameter $a$. We also
compare our results with those obtained within the perturbative
method and we discuss the differences between them. }


\arxivnumber{}

\begin{document}

\maketitle \flushbottom

\section{Introduction}

General relativity (GR) is well-tested in the weak-field regime,
whereas the strong-field regime remains essentially unexplored and
unconstrained. Namely in the strong regime, GR is expected to break
down and give way to a more complete theory of gravitation. Indeed,
it is well-known that the quantum corrections in the strong field
regime give rise to a  modification of Einstein gravity -- the
renormalization at one loop demands that the Einstein-Hilbert action
be supplemented with higher order terms \cite{BD1982}. On the other
hand, the attempts to construct a unified theory of the
interactions, naturally lead to scalar-tensor type generalizations
of General Relativity and theories of gravity with Lagrangians
containing various kinds of curvature corrections to the usual
Einstein-Hilbert Lagrangian \cite{CFPS2012}.

Besides the theoretical reasons, there are  well-known observational
facts that force us to go beyond the original Einstein theory and to
study modified gravity theories. One of the most important
discoveries in physics in the last two decades was the accelerated
expansion of the Universe. The cause for accelerated expansion of
the Universe is a mystery at present, and it is a great challenge
for physics to solve this problem. There are, however, two general
hypothesis for explaining the acceleration of the Universe.
According to the first hypothesis, there exists a new kind of
matter, called dark energy, which governs the accelerated expansion
of the Universe \cite{R1}. Within the framework of this hypothesis
the dark energy constitutes $73\%$ of the total energy content of
the Universe and it exhibits some rather unusual and strange
properties such as negative pressure to density ratio.

The second hypothesis for the explanation of the accelerated
expansion of the universe is that  general relativity (i.e. Einstein
equations) should be modified. In other words, instead of
attributing the accelerated expansion to unknown constituents of the
Universe with rather unusual and strange properties, one can
attribute it to our lack of understanding of gravity. In particular,
one may consider that the accelerated expansion indicates the break
down of general relativity at cosmological and even at astrophysical
scales. Ones of the  most popular generalized gravitational theories
which can give a possible explanation of the present cosmological
observations are the  $f(R)$ theories
\cite{SF2010},\cite{FT2010},\cite{ON2011}.

The $f(R)$ theories, as it is well-known, are modifications of GR in
which the usual Einstein-Hilbert Lagrangian is replaced with an
appropriately chosen function of the scalar curvature $R$. By their
very construction, the $f(R)$ theories inevitably contain
dimensionful parameters which, at present, should be found or
constrained by the experiments and the observations. In general,
these  parameters could determine rather different scales (e.g. from
scales typical for the compact objects to cosmological scales) and
can be responsible for the strong and weak field regime of the
theory.

Neutron stars are natural laboratories for  investigating  the
strong-field regime of gravity and for testing the alternative
gravitational theories.  That is why neutron stars in $f(R)$
theories are currently an active field of research and they were
studied by many authors \cite{Cooney09}--\cite{Astashenok2014}. The
study of  neutron star structure in $f(R)$ theories also might allow
us to impose constraints on the  parameters responsible for the
strong field regime. A serious problem in using the neutron stars
for testing the alternative theories and imposing constraints on
their parameters is the uncertainty in the equation of state (EOS).
The EOS uncertainty does not allow for accurate parameter
constraints when the deviations from GR are of the same order as the
deviations induced by different EOS. Nevertheless, at present we
have some observational constraints on mass-radius relation for
neutron stars which can be used in some extent to eventually
constrain the parameters responsible for the strong field regime of
the alternative theories.

A drawback of most of the  works on neutron stars in $f(R)$ gravity
is the method of investigation adopted in them. They use a
perturbative scheme  in order to study the neutron stars in $f(R)$
theories.  More precisely they consider $f(R)$ theories as a
perturbation to GR.  In our opinion the use of a perturbative method
to investigate the strong field regime in $f(R)$ theories is not
completely justified, and might lead to unphysical results. It is
well known that even if in the weak field regime modified gravity is
indistinguishable from general relativity, nonlinear phenomena may
appear in the strong field regime with serious consequences for the
structure and the properties of the compact objects
\cite{Damour1993}–-\cite{DYSK13}.

Another drawback in most of the current studies of neutron stars in
$f(R)$ theories is the fact that the interior and exterior problems
are not solved simultaneously as the self-consistent approach
requires. Instead, the exterior solution is imposed to be the
Schwarzschild one which is equivalent to freeze the scalar degree of
freedom outside the star in contradiction  with the field equations.
In this way the self-consistency of the whole problem is violated
and this may  lead (and leads) to artificial effects on the neutron
star structure. It is also worth mentioning that  some papers
consider strong field regime parameters with wrong sign which means
that the conditions for absence of tachyonic instabilities are not
satisfied leading to new but non-physical branches of neutron star
solutions. For example, such non-physical solutions were obtained
within the framework of the $R$-squared gravity defined by $f(R)=R +
a R^2$. In addition to the physical solutions with $a>0$ in
\cite{Arapoglu11}, new but non-physical solutions were also obtained
there by considering the unphysical sector with $a<0$ plagued with a
tachyonic instability and with a seemingly wrong asymptotic
expansion.

The aim of this paper is to investigate non-perturbatively and
self-consistently  the neutron stars in  one of the best known
examples of $f(R)$ theories, namely  the $R$-squared gravity $f(R)=R
+ a R^2$ by simultaneously  solving the interior and the exterior
problem. We also comment on the possible constraints on the
parameter $a$ that can be imposed if the current observations of the
neutron star mass and radius are taken into account.

  The non-perturbative and
self-consistent  re-examination of neutron stars in $f(R)$ theories
is important for several reasons. The comparison between the
perturbative methods and the non-perturbative one is very
instructive for the future studies. The constraints on the
parameters of the $f(R)$ theories obtained via the perturbative
methods may be  misleading in confronting the theory with the
experiments and observations. The non-perturbative method can cover
much wider ranges of the parameters and the parameter constrains
based on it are much more reliable.  Moreover, the perturbative
methods predicts the possible existence of qualitatively new stable
neutron star branches in comparison with GR -- a rather drastic
prediction in its own. So, non-perturbative confirmation or
rejection of the possible existence of new stable neutron star
branches is needed.

As we shall see below the non-perturbative method gives results that
are  different from the ones obtained by the perturbative method.

The paper is organized as follows. In Section 2 we give the
basic framework of the problem and the reduced field equations that
will be solved numerically. In Section 3 we present and discuss the
results for neutron stars with realistic equations of state in
non-perturbative $R$-squared gravity. We also compare our results
with those obtained via the perturbative approach. We end the paper
with discussion and conclusions.

\section{Basic equations}

The action of the $f(R)$ theories is given by

\begin{eqnarray}\label{A}
S= \frac{1}{16\pi G} \int d^4x \sqrt{-g} f(R) + S_{\rm
matter}(g_{\mu\nu}, \chi),
\end{eqnarray}
where $R$ is the scalar curvature with respect to the spacetime
metric $g_{\mu\nu}$ and $S_{\rm matter}$ is the action of the matter
fields denoted by $\chi$. The viable $f(R)$ theories have to be free
of tachyonic instabilities and the appearance of ghosts which
require \cite{SF2010},\cite{FT2010}

\begin{eqnarray}
\frac{d^2f}{dR^2}\ge 0,  \;\;\; \frac{df}{dR}>0,
\end{eqnarray}
respectively.

It is well-known that the $f(R)$ theories are equivalent to the
Brans-Dicke scalar-tensor theory with $\omega_{BD}=0$ and with a
potential for the scalar field. This can be easily demonstrated by
considering a new field $\psi$ and the dynamically equivalent action

\begin{eqnarray}\label{A1}
S= \frac{1}{16\pi G} \int d^4x \sqrt{-g}\left[f(\psi) +
f^{\prime}(\psi)(R-\psi)\right] + S_{\rm matter}(g_{\mu\nu}, \chi).
\end{eqnarray}

Varying then with respect to $\psi$ one obtains
$f^{\prime\prime}(\psi)(R-\psi)=0$ which, provided that
$f^{\prime\prime}(\psi)\ne 0$, gives $\psi=R$. This substituted back
in (\ref{A1}) indeed recovers (\ref{A}). Introducing the new field
$\Phi=f^{\prime}(\psi)$ and defining the potential $U(\Phi)$ via

\begin{eqnarray}
U(\Phi)= \psi(\Phi)f^{\prime}(\Psi(\Phi)) - f(\psi(\Phi)),
\end{eqnarray}
our action (\ref{A1}) takes exactly the form of the action of
Brans-Dicke theory in Jordan frame with a potential for the scalar
field, namely

\begin{eqnarray}
S=\frac{1}{16\pi G} \int d^4x \sqrt{-g}\left[\Phi R - U(\Phi)\right]
+ S_{\rm matter}(g_{\mu\nu}, \chi).
\end{eqnarray}

Especially, in the case of $R$-squared gravity ($f(R)= R + aR^2$),
the Brans-Dicke potential is given by

\begin{eqnarray}
U(\Phi)=\frac{1}{4a}(\Phi - 1)^2,
\end{eqnarray}
which corresponds to a massive scalar field with a mass
$m_{\Phi}=\frac{1}{\sqrt{6a}}$. Here we consider only non-negative
values for the parameter $a$ which obey the condition
$\frac{d^2f}{dR^2}\ge 0$.

From a mathematical and numerical point of view, as in the general
scalar-tensor theories, it is more convenient to study the field
equations in the so-called Einstein frame. The Einstein frame is
defined by introducing the new scalar field $\varphi$ and the new
metric $g^{*}_{\mu\nu}$  given by

\begin{eqnarray}
&&\varphi =\frac{\sqrt{3}}{2}\ln\Phi, \\
&&g^{*}_{\mu\nu} = \Phi g_{\mu\nu}= A^{-2}(\varphi)g_{\mu\nu},
\end{eqnarray}
with $A^2(\varphi)=\Phi^{-1}(\varphi)=
e^{-\frac{2}{\sqrt{3}}\varphi}$. The Einstein frame action then
takes the form

\begin{eqnarray}\label{EFA}
S=\frac{1}{16\pi G} \int d^4x \sqrt{-g^{*}}\left[ R^{*} - 2
g^{*\mu\nu}\partial_{\mu}\varphi \partial_{\nu}\varphi - V(\varphi)
\right] + S_{\rm
matter}(e^{-\frac{2}{\sqrt{3}}\varphi}g^{*}_{\mu\nu},\chi),
\end{eqnarray}
where $R^{*}$ is the Ricci scalar curvature with respect to the
Einstein frame metric $g^{*}_{\mu\nu}$ and
$V(\varphi)=A^4(\varphi)U(\Phi(\varphi))$. For the $R$-squared
gravity the explicit form of the potential in the  Einstein frame is

\begin{eqnarray}
V(\varphi)= \frac{1}{4a}
\left(1-e^{-\frac{2\varphi}{\sqrt{3}}}\right)^2 . \label{eq:Potential}
\end{eqnarray}

The expense for simplifying the action in Einstein frame (and the
field equations as a consequence) is the appearance of direct
interaction between the matter fields and the scalar field $\varphi$
in this frame. Taking variation with respect to the metric
$g^{*}_{\mu\nu}$ and the scalar field $\varphi$, we find the field
equations in the Einstein frame

\begin{eqnarray}
&&G^{*}_{\mu\nu}= 8\pi G T^{*}_{\mu\nu} +
2\partial_{\mu}\varphi\partial_{\nu}\varphi - g^{*}_{\mu\nu}
{g^{*}}^{\alpha\beta}\partial_{\alpha}\varphi\partial_{\beta}\varphi
- \frac{1}{2}V(\varphi)g^{*}_{\mu\nu}, \\
&&\nabla^{*}_{\mu}\nabla^{*\mu}\varphi -
\frac{1}{4}\frac{dV(\varphi)}{d\varphi}= - 4\pi G \alpha(\varphi)
T^{*},
\end{eqnarray}
where

\begin{eqnarray}
\alpha(\varphi)= \frac{d\ln
A(\varphi)}{d\varphi}=-\frac{1}{\sqrt{3}}.
\end{eqnarray}

The Einstein frame energy-momentum tensor $T^{*}_{\mu\nu}$ is
related to the Jordan frame one $T_{\mu\nu}$ via
$T^{*}_{\mu\nu}=A^{2}(\varphi)T_{\mu\nu}$. In the case of a perfect
fluid, the energy density, the pressure and the 4-velocity in the
two frames are related via the formulae

\begin{eqnarray}\label{EJFPF}
\rho_{*}=A^{4}(\varphi)\rho, \;\; p_{*}=A^{4}(\varphi) p, \;\;
u^{*}_{\mu}=A^{-1}(\varphi) u_{\mu}.
\end{eqnarray}

The contracted Bianchi identities give the following conservation
law for the Einstein frame energy-momentum tensor

\begin{eqnarray}
\nabla^{*}_{\mu}T^{*\mu}_{\;\;\nu}= \alpha(\varphi)
T^{*}\nabla^{*}_{\nu}\varphi .
\end{eqnarray}

The next step is to consider a static and spherically symmetric
spacetime described by  the Einstein frame metric

\begin{eqnarray}
ds^2_{*}= - e^{2\phi(r)}dt^2 + e^{2\Lambda(r)}dr^2 + r^2(d\theta^2 +
\sin^2\theta d\vartheta^2 ).
\end{eqnarray}

Since the purpose of the present paper is to study the structure of
neutron stars in $f(R)$ gravity, we consider the matter source to be
a perfect fluid. We also require the perfect fluid and the scalar
field to respect the staticity and the spherical symmetry. With
these conditions imposed, the dimensionally  reduced field equations
are

\begin{eqnarray}
&&\frac{1}{r^2}\frac{d}{dr}\left[r(1- e^{-2\Lambda})\right]= 8\pi G
A^4(\varphi) \rho + e^{-2\Lambda}\left(\frac{d\varphi}{dr}\right)^2
+ \frac{1}{2} V(\varphi), \label{eq:FieldEq1} \\
&&\frac{2}{r}e^{-2\Lambda} \frac{d\phi}{dr} - \frac{1}{r^2}(1-
e^{-2\Lambda})= 8\pi G A^4(\varphi) p +
e^{-2\Lambda}\left(\frac{d\varphi}{dr}\right)^2 - \frac{1}{2}
V(\varphi),\label{eq:FieldEq2}\\
&&\frac{d^2\varphi}{dr^2} + \left(\frac{d\phi}{dr} -
\frac{d\Lambda}{dr} + \frac{2}{r} \right)\frac{d\varphi}{dr}= 4\pi G
\alpha(\varphi)A^4(\varphi)(\rho-3p)e^{2\Lambda} + \frac{1}{4}
\frac{dV(\varphi)}{d\varphi} e^{2\Lambda}, \label{eq:FieldEq3}\\
&&\frac{dp}{dr}= - (\rho + p) \left(\frac{d\phi}{dr} +
\alpha(\varphi)\frac{d\varphi}{dr} \right), \label{eq:FieldEq4}
\end{eqnarray}
and they describe the interior structure, i.e. the spacetime metric,
the energy density, pressure and scalar field inside the neutron
star. Note that in the above system we  used (\ref{EJFPF}) and we
substituted $\rho_{*}$ and $p_{*}$ with $A^{4}(\varphi)\rho$ and
$A^{4}(\varphi) p$, respectively.

The equations describing the spacetime metric and the scalar field
outside the neutron star are obtained from the above system by
formally putting $\rho=p=0$. In addition to our systems of
differential equations for the interior and the exterior of the
neutron star, we should give the equation of state (EOS) for the
neutron star matter $p=p(\rho)$ and  impose the boundary conditions.

As we have already discussed in the introduction, we solve the
interior and the exterior problem simultaneously with the following
natural Einstein frame boundary conditions in the center of the star

\begin{eqnarray}
\rho(0)=\rho_{c},\;\;\; \;\Lambda(0)=0,\;\;\;\;
\frac{d\varphi}{dr}(0)=0,\label{eq:BC1}
\end{eqnarray}
and at infinity

\begin{eqnarray}\label{BCINF}
\lim_{r\to \infty}\phi(r)=0, \;\;\;\; \lim_{r\to \infty}\varphi
(r)=0. \label{eq:BC2}
\end{eqnarray}

The coordinate radius $r_S$ of the star is determined by the
condition

\begin{eqnarray}\label{NSR}
p(r_S)=0.
\end{eqnarray}

Some comments on the boundary conditions are in order. The condition
$\frac{d\varphi}{dr}(0)=0$ ensures the regularity of the scalar
field $\varphi$, and in turn the regularity of $\Phi$, at the center
$r=0$. The regularity of the Einstein frame geometry at the center
requires $\Lambda(0)=0$. Since the Jordan and the Einstein frame
metrics are conformally related (via nonsingular conformal factor)
this condition also insures the regularity of the Jordan frame
geometry at the center of the star. The boundary conditions at
infinity are related to the fact that we consider neutron stars in
asymptotically flat spacetime\footnote{The assumption for
asymptotically flat spacetime is completely justified for local
astrophysical systems of size of 1 AU or less, compared to the
cosmological scales on the order of $10^{26} m$. }. The asymptotic
flatness requires $\lim_{r\to\infty} V(\varphi(r))=0$ which gives
$\lim_{r\to\infty}\varphi(r)=0$ or equivalently
$\lim_{r\to\infty}\Phi(r)=1$. Note that the conditions (\ref{BCINF})
ensure the asymptotic flatness in both Einstein and Jordan frame.

Although the coordinate radius of the star is determined via
(\ref{NSR}),  the physical radius of the star as measured in the
physical Jordan frame is given by

\begin{eqnarray}
R_{S}= A[\varphi(r_S)] r_S.
\end{eqnarray}

Concerning the mass of the neutron star, it can be found from the
asymptotic expansion of the physical Jordan frame metric. However,
especially for the case of $R$-squared gravity, the scalar field
$\Phi$ (respectively $\varphi$) has a finite range, i.e. it drops
off exponentially at infinity and this shows that the Jordan and the
Einstein frame masses coincide.

In the numerical results, presented in the next section, we  use the
dimensionless parameter

\begin{eqnarray}
a \to \frac{a}{R^2_{0}},
\end{eqnarray}
where $R_{0}=1.47664$ km which corresponds to one solar mass.

\section{Numerical results}

The field equations \eqref{eq:FieldEq1}--\eqref{eq:FieldEq4}
together with the boundary conditions \eqref{eq:BC1} and
\eqref{eq:BC2} are solved numerically using a shooting method.
Additional complications come from the fact that the presence of a
nontrivial potential of the form \eqref{eq:Potential} makes the
system of differential equations stiff, with increasing stiffness as
$a$ decreases. This requires refinement of the numerical algorithm
and a close control of the shooting procedure.

Below we present in detail the obtained results. We use four
realistic EOS with distinct properties in order to show the possible
deviations from general relativity more thoroughly. The equations of
state are SLy4 \cite{SLy4}, APR4 \cite{APR4}, FPS \cite{FPS} and L
\cite{EOSL}\footnote{For EOS APR4 we used the piecewise polytropic
approximation given in \cite{Read2009}.}. EOS SLy4 and APR4 are both
modern realistic equations of state that fulfill all of the
observational constrains on the neutron star mass and radius
\cite{Lattimer2012}--\cite{Demorest2010}. EOS FPS on the other hand
is softer and its maximum mass does not reach the two solar mass
barrier \cite{Antoniadis2013,Demorest2010}. Nevertheless we examined
it because, as we will show below, $f(R)$ gravity could alter
considerably the maximum mass of neutron stars and eventually
reconcile an EOS with the observations. EOS L is one of the stiffest
proposed realistic EOS and even though it leads to somewhat larger
radii than the current observational constraints, we will consider
it as a limiting case in stiffness.

\begin{figure}[]
\centering
\includegraphics[width=0.48\textwidth]{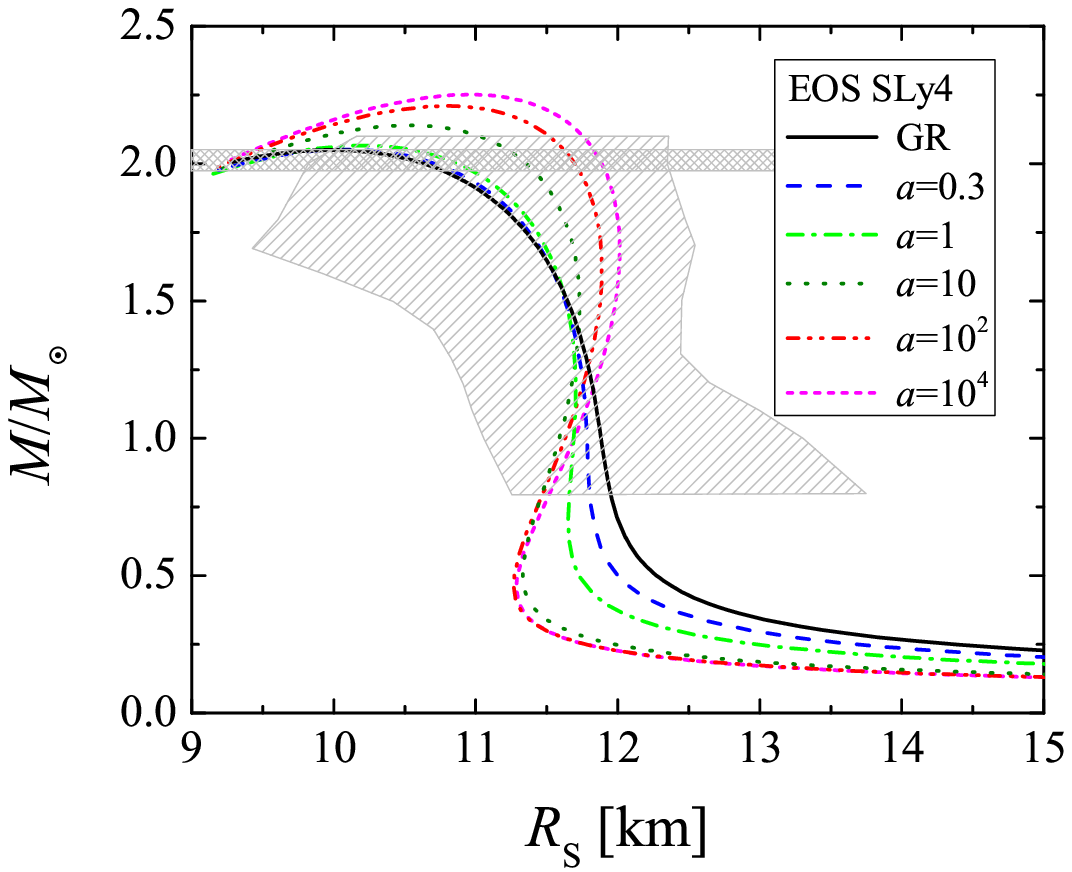}
\includegraphics[width=0.48\textwidth]{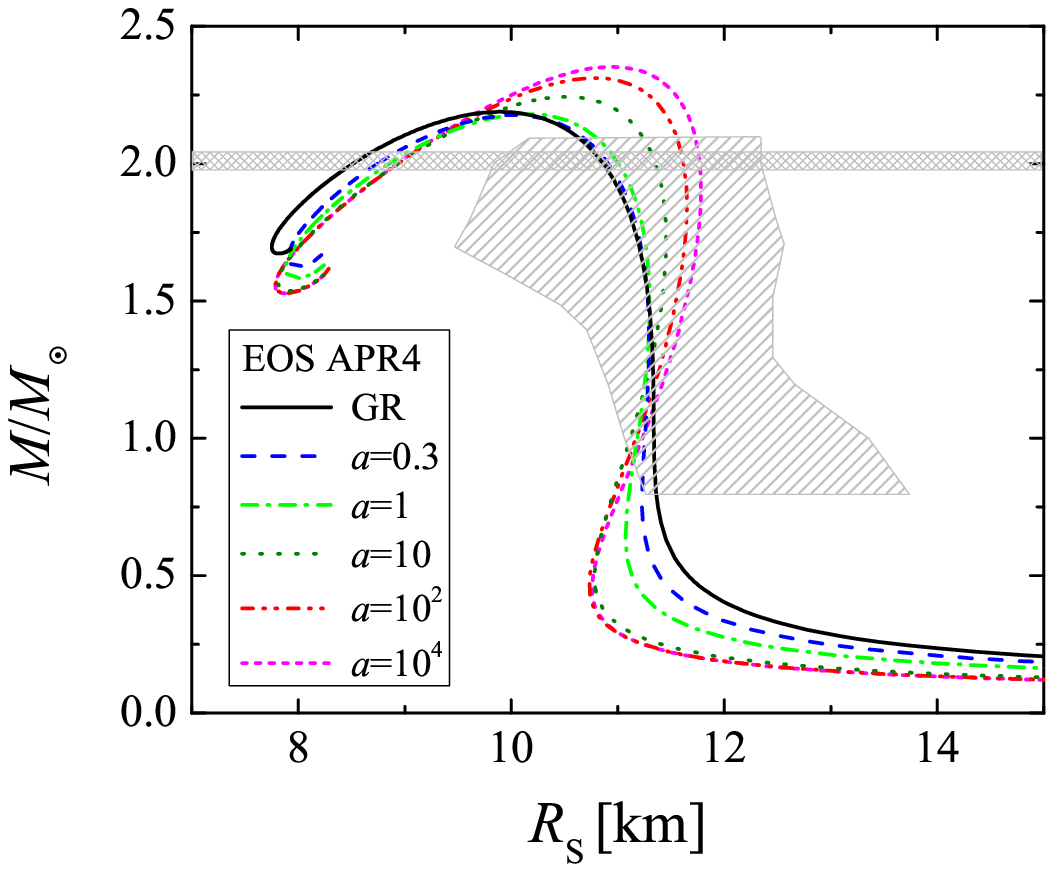}
\caption{The mass of radius relation for EOS SLy4 (left panel) and
APR4 (right panel). Different styles and colors of the curves
correspond to different values of the parameter  $a$. The current
observational constrains are shown as shaded regions.}
\label{Fig:M(R)_SLy_APR}
\end{figure}

\begin{figure}[]
\centering
\includegraphics[width=0.48\textwidth]{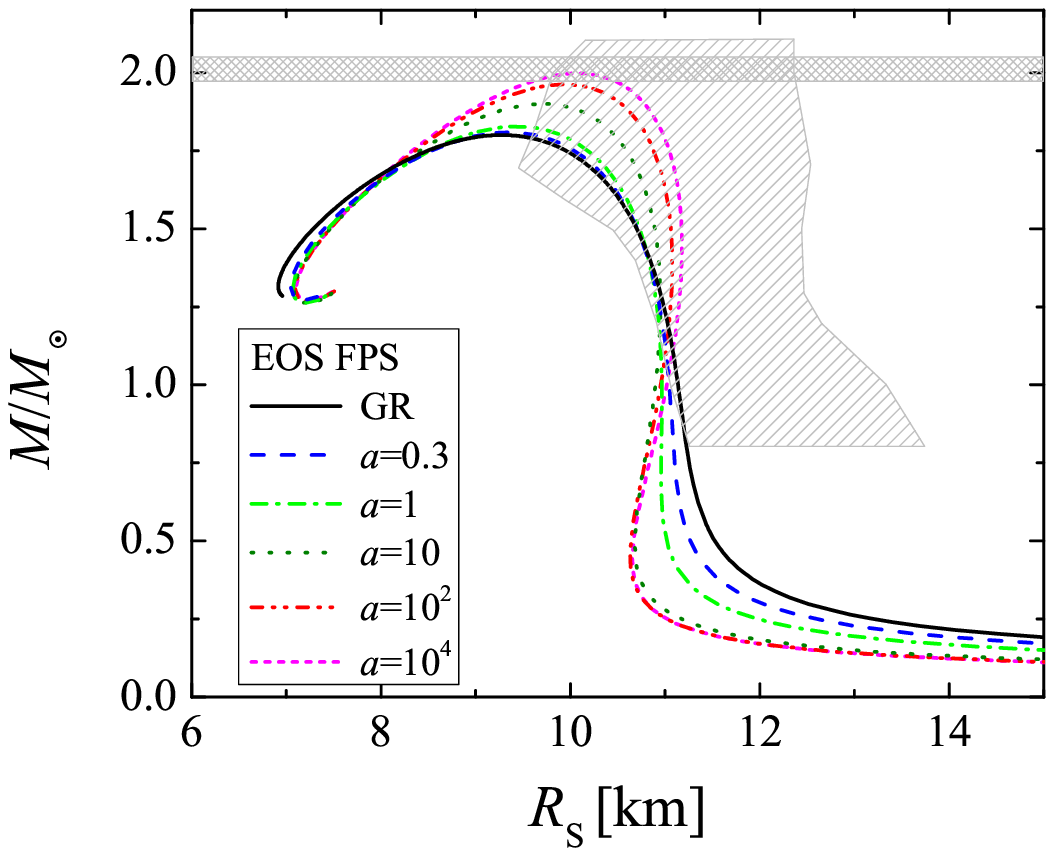}
\includegraphics[width=0.48\textwidth]{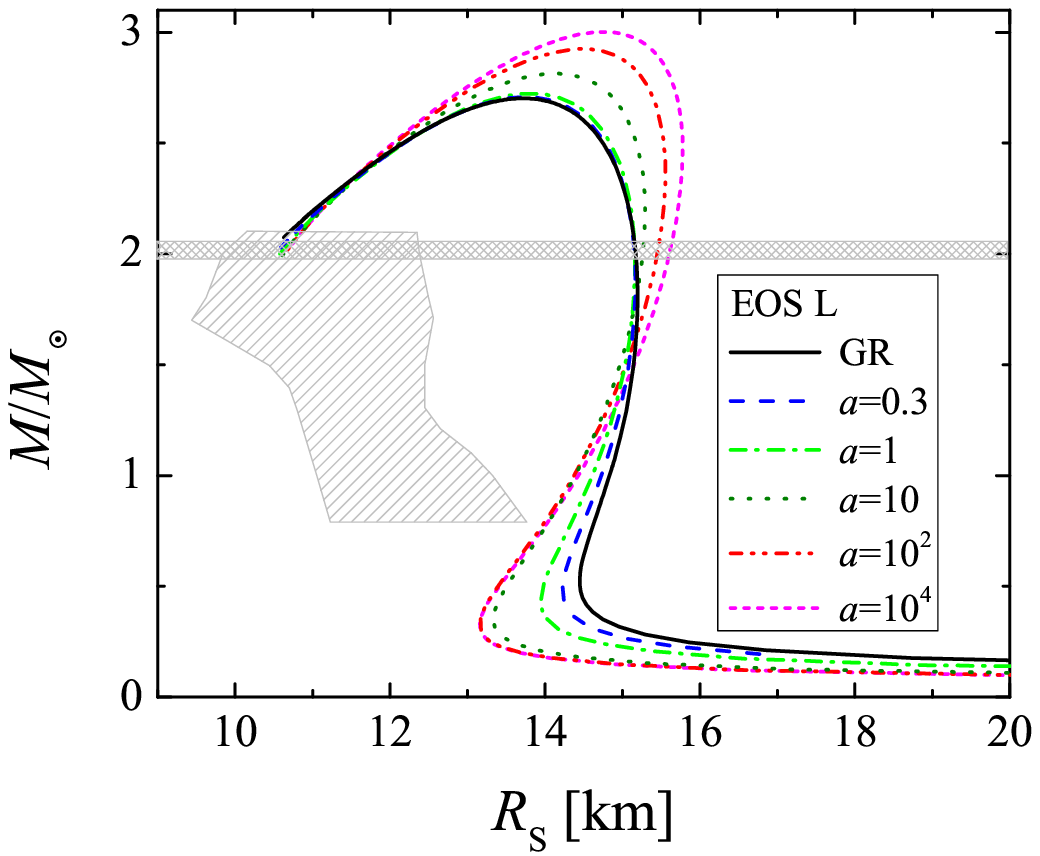}
\caption{The mass of radius relation for EOS FPS (left panel) and L
(right panel). Different styles and colors of the curves correspond
to different values of the parameter  $a$. The current observational
constrains are shown as shaded regions.} \label{Fig:M(R)_FPS_L}
\end{figure}

The mass of radius relations for the four realistic EOS are given in
Figs. \ref{Fig:M(R)_SLy_APR} and \ref{Fig:M(R)_FPS_L}. The current
observational constraints on the neutron star mass and radius, given
in \cite{Steiner2010,Antoniadis2013}, are shown as shaded regions on
the graphs. Lines with different styles and colors in every figure
correspond to different values of the parameter $a$ ranging from
$a=0.3$ to $a=10^4$\footnote{We studied systematically the whole
range from $a=0.02$ to $a=10^5$, but we use a narrower range of $a$
for a better representation of the data. The models with $a<0.3$ and
$a>10^4$ are very close to the cases of $a=0.3$ and $a=10^4$
respectively.}. As one can see, for all of the EOS, $a=0.3$ leads to
models that are almost indistinguishable from general relativity
especially for masses above one solar mass and for $a<0.3$ the
neutron star solutions get closer and closer to the general
relativistic ones. On the other hand when $a \rightarrow \infty$ the
neutron star mass and radius saturates to certain values for a fixed
central energy density and EOS, and $a=10^4$ gives us nearly the
maximum possible deviation from the pure Einstein theory. The reason
for this behavior is the following. The increase of $a$ is
equivalent to decreasing the scalar-field mass: when $a\rightarrow
\infty$ the mass of the scalar field vanishes and when $a\rightarrow
0$ the mass goes to infinity. Loosely speaking the nonzero mass
suppresses the scalar field exponentially, and in general larger
masses correspond to smaller values of the scalar field. That is why
models with smaller $a$ are closer to general relativity, which is
recovered in the limit $a\to 0$. When $a$ increases, the mass of the
scalar field decreases and thus the scalar field can reach larger
values. Consequently the deviations from general relativity also
increase. The case $a \rightarrow \infty$ corresponds to
$\omega_{BD}=0$ Brans-Dicke scalar-tensor theory with zero potential
of the scalar field and represents the maximum possible difference
between the considered $f(R)$ theory of gravity and general
relativity.

A common feature for all of the realistic EOS is the following --
for larger neutron star masses the presence of nontrivial scalar
field leads to an increase of the neutron star radius. For smaller
neutron star masses on the other hand the radius decreases. This
behavior is qualitatively different from the perturbative results
presented in \cite{Arapoglu11}, where for large masses the radius
decreases and for small masses it increases compared to Einstein
theory.

For all of the considered equations of state the maximum mass
reaches up to approximately $10\%$ larger values in the limit
$a\rightarrow \infty$ compared with general relativity. This fact
can help us for example to reconcile EOS FPS with the observation.
In the general relativistic case its maximum mass is significantly
below two solar masses, but for large values of the parameter $a$
the two solar mass barrier is reached. A similar observation can be
made for EOS SLy4 that barely reaches two solar masses in the pure
Einstein theory, but when $a$ increases the maximum mass also
increases and the mass constraint is easily satisfied. On the other
hand the maximum mass of the neutron star sequences can decrease for
small values of $a$ (typically for $a<2$). But this decrease of the
mass is very small -- it does not exceed $1\%$ for all of the
realistic EOS.

{\it As a whole the results in the present paper lead us to the idea
that the differences between the $R$-squared gravity  and general
relativity are comparable with the uncertainties in the nuclear
matter equations of state. That is why the current observations of
the neutron star masses and radii alone can not put constraints on
the value of the parameters $a$, unless the equation of state is
better constrained in the future.}

\begin{figure}[]
\centering
\includegraphics[width=0.48\textwidth]{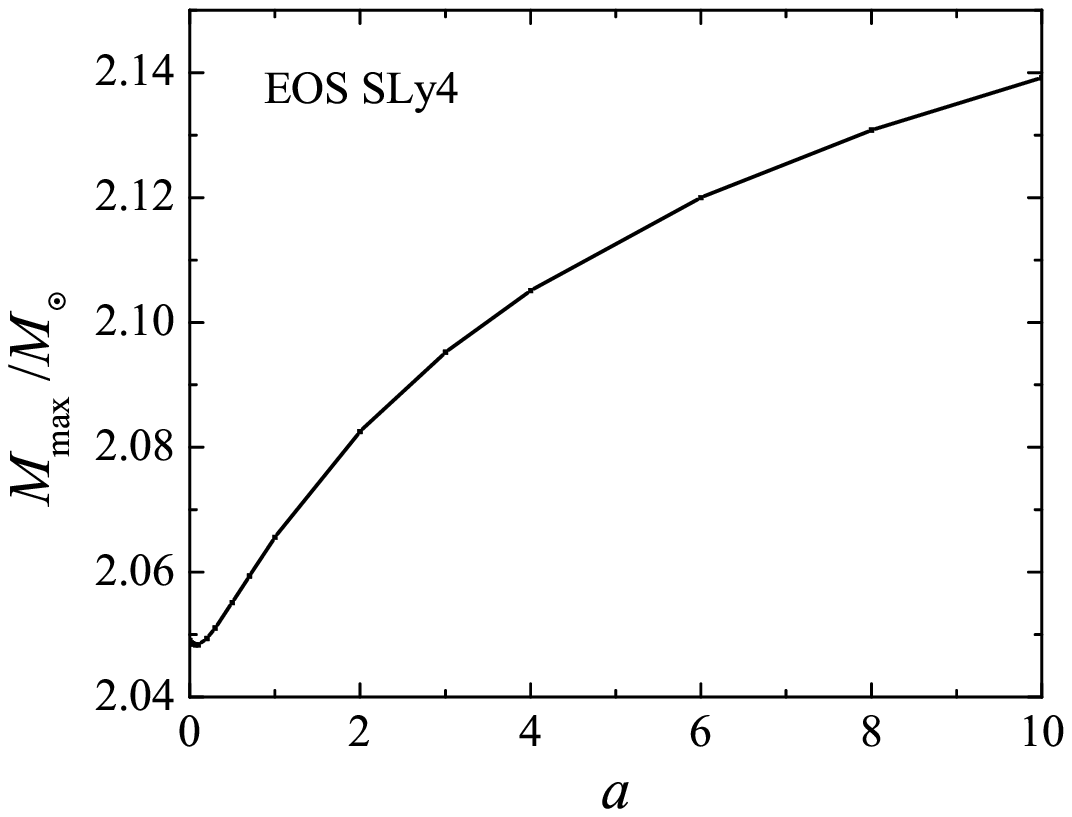}
\includegraphics[width=0.48\textwidth]{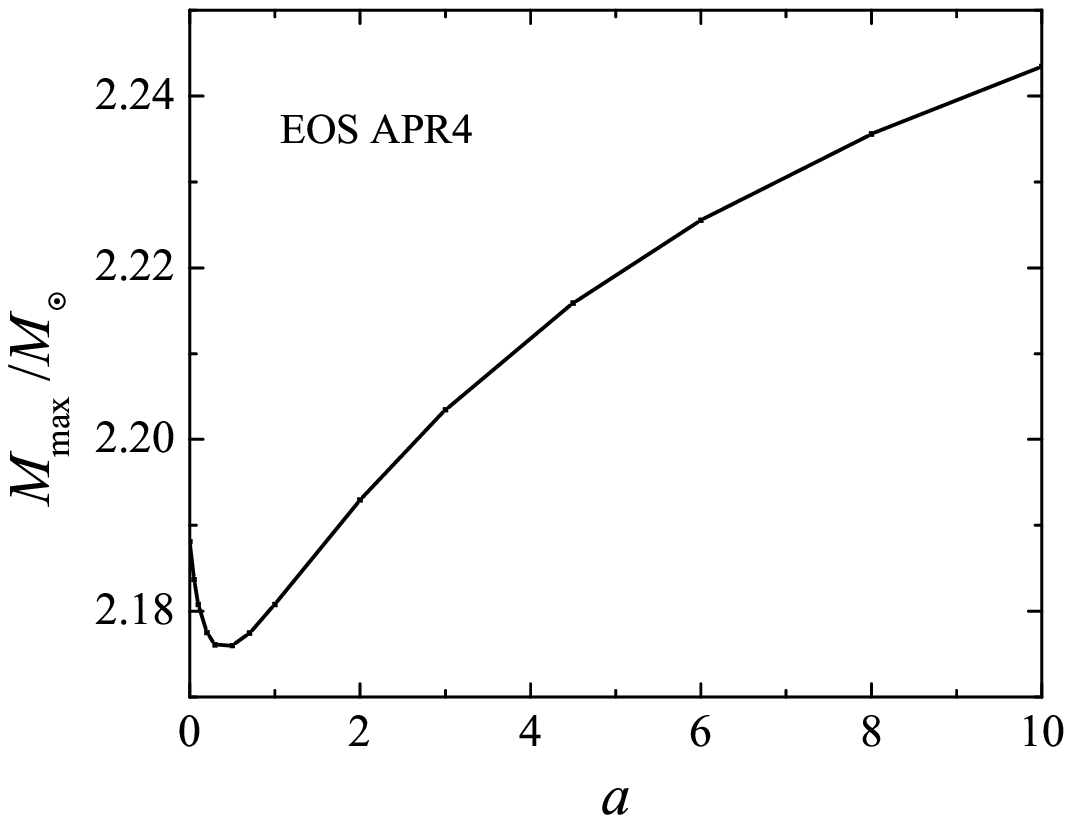}
\caption{The maximum mass as a function of the parameter $a$ for EOS
SLy4 (left panel) and APR4 (right panel). In the limit $a\rightarrow
\infty$, $M_{max} = 2.25 M_\odot$ for EOS SLy4 and $M_{max} = 2.36
M_\odot$ for EOS APR4.} \label{Fig:Mmax_SLy4_APR}
\end{figure}

It is interesting to examine more thoroughly how does the maximum
mass changes when we vary the parameter $a$ and to compare our
results with the ones obtained by the perturbative approach  in
\cite{Arapoglu11}. In Fig. \ref{Fig:Mmax_SLy4_APR} we have plotted
the maximum mass $M_{\rm max}$ as a function of $a$ for EOS SLy4 and
APR4. In the figures the general relativistic case is presented as
$a=0$. The behavior of the plotted dependences is characteristic for
all of the studied EOS. When $a \rightarrow 0$ the maximum mass
tends to the corresponding value in general relativity for a
particular EOS. With the increase of $a$, $M_{\rm max}$ first
decreases and after reaching a minimum it starts to increase
monotonically. The maximum value of $M_{\rm max}$ is reached for
$a\rightarrow \infty$ and it is up to approximately $10\%$ larger
than the pure Einstein theory depending on the EOS. The minimum is
typically reached for $a<1$ and $M_{\rm max}$ deviates less than
$1\%$ compared with pure general relativity. For the considered
realistic equations of state the radius of the corresponding maximum
mass models $R_{\rm min}$ is a monotonically increasing function of
$a$.

The decrease of $M_{\rm max}$ for small values of $a$ was also
observed in \cite{Arapoglu11}, but the subsequent increase of the
maximum mass was not reported there for any of the realistic EOS. In
general there are noticeable quantitative differences between our
results and those of \cite{Arapoglu11}. The most important is that
the decrease of the maximum mass observed in \cite{Arapoglu11} is
much stronger compared to the non-perturbative approach. As a
consequence constraints on $a$ were obtained in \cite{Arapoglu11}
after a comparison with observation (more precisely they require
that the maximum mass of the modern realistic equations of state
should not fall below approximately two solar masses). Such
constraints are obviously not possible in the non-perturbative
approach where the decrease of the mass is very small and the
deviations from general relativity are comparable with the
deviations coming from the use of different modern realistic
equations of state.

\begin{figure}[]
\centering
\includegraphics[width=0.48\textwidth]{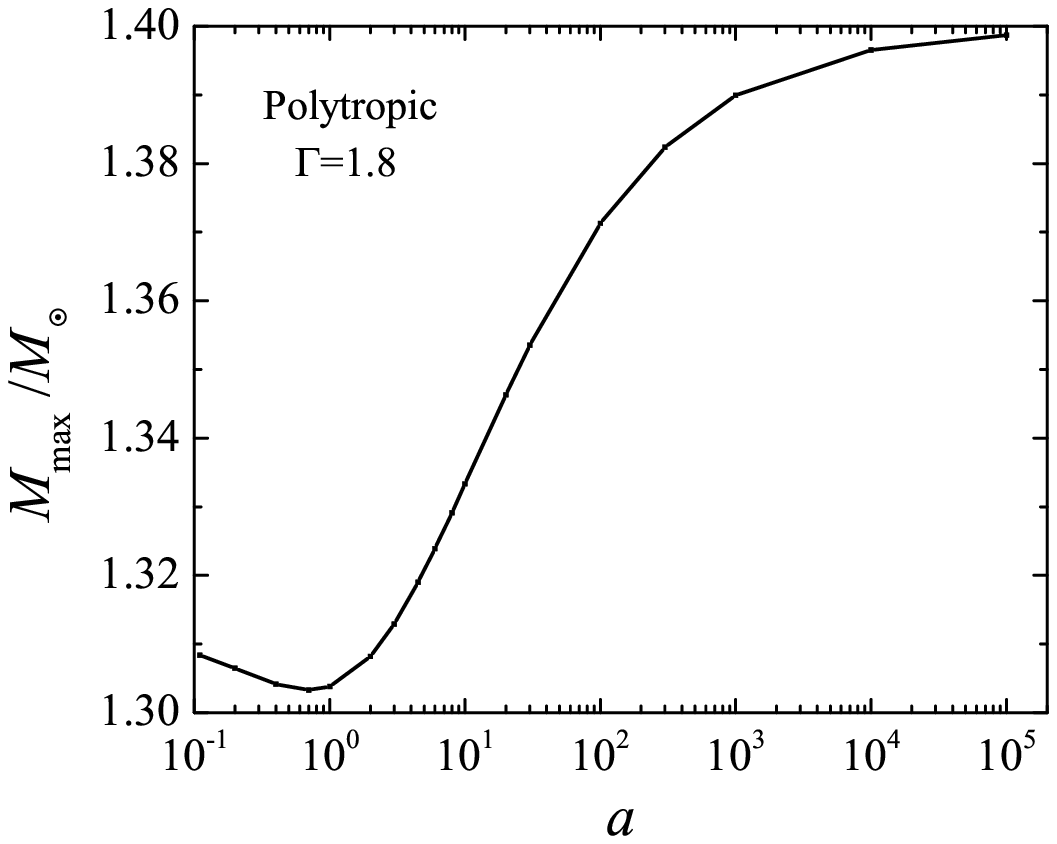}
\includegraphics[width=0.48\textwidth]{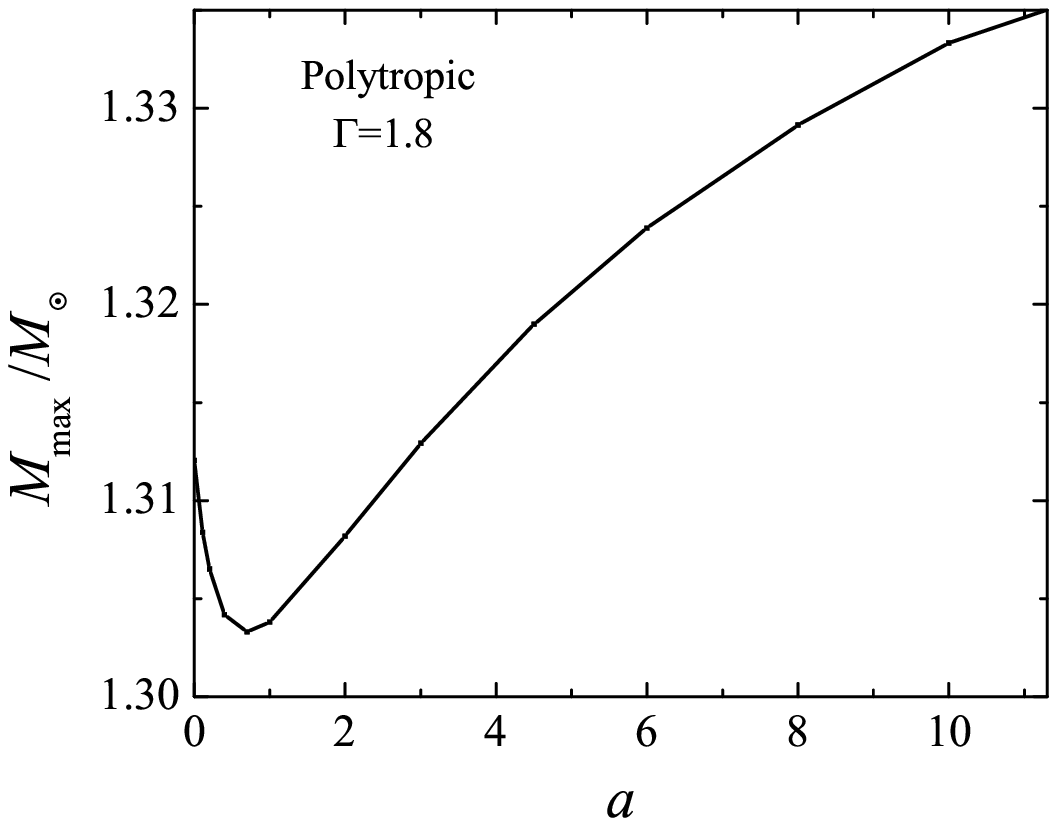}
\caption{The maximum mass as a function the parameter $a$ for polytropic EOS with index $\Gamma=1.8$.
The right figure is a magnification in a non-logarithmic scale. }
\label{Fig:Mmax_poly}
\end{figure}

\begin{figure}[]
\centering
\includegraphics[width=0.48\textwidth]{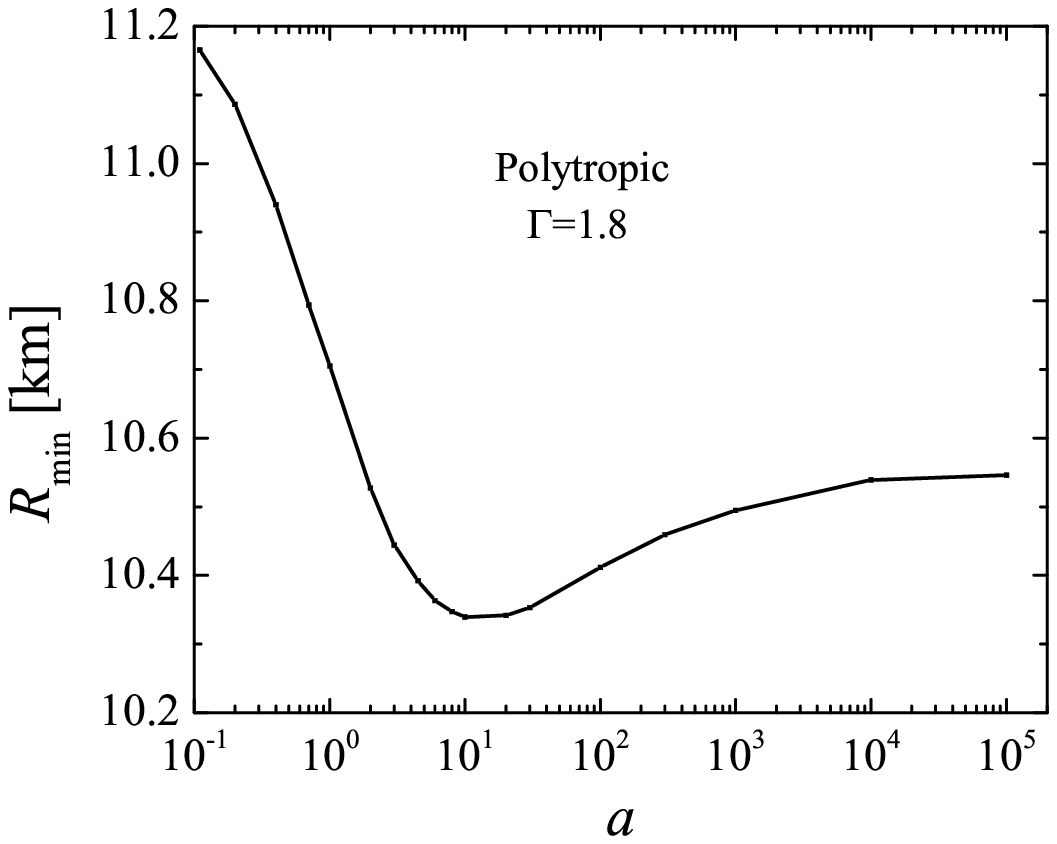}
\includegraphics[width=0.48\textwidth]{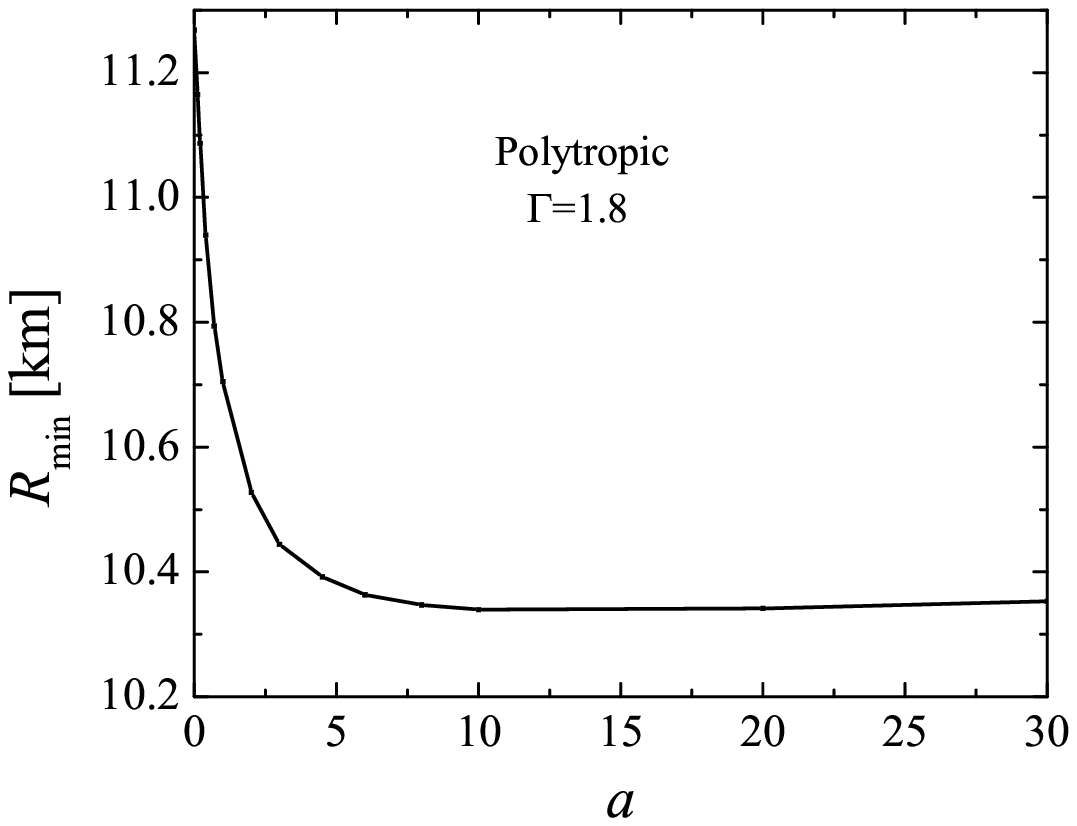}
\caption{The minimum radius as a function the parameter $a$ for polytropic EOS with index $\Gamma=1.8$.
The right figure is a magnification in a non-logarithmic scale.}
\label{Fig:Rmax_poly}
\end{figure}

In order to be more precise we also made a systematic comparison
with the results for a specific polytropic equation of state given
in \cite{Arapoglu11}. The value of the polytropic index used there
is $\Gamma=9/5$. The mass $M_{\rm max}$ and radius $R_{\rm min}$ of
the maximum mass models as a function of $a$ are presented in Figs.
\ref{Fig:Mmax_poly} and \ref{Fig:Rmax_poly} for this polytropic EOS.
The qualitative behavior of $M_{\rm max}$ and $R_{\rm min}$ is
similar to the results in \cite{Arapoglu11} -- when we increase $a$
the values of $M_{\rm max}$ and $R_{\rm min}$ first decrease and
after reaching a minimum they start to increase. Also the minima of
$M_{\rm max}$ is at similar values of the parameter $a$ compared to
the results in \cite{Arapoglu11}. But the quantitative differences
with the corresponding figures in \cite{Arapoglu11} are significant.
For example the minimum of $M_{\rm max}$ is considerably deeper in
\cite{Arapoglu11} compared to our results. The minimum of $R_{\rm
min}$ on the other hand is located at much larger values of $a$ in
our case and it reaches considerably smaller values of $R_{\rm min}$
compared to the perturbative approach.

It is also worth confronting the non-perturbative and
self-consistent approach with the perturbative approach in their
description of the local internal structure of the neutron stars.
Within the perturbative and non-self-consistent  approach it was
found in \cite{OGPR2013} that for the equation of state SLy there
are regions inside the star where the local Jordan frame mass
decreases with the radial coordinate in a contra-intuitive way. For
the same equation of state and the same value of the parameter $a$
no such behaviour is observed in the non-perturbative approach. The
same holds also for the other EOS.

Another fact showing  that the perturbative approach is not
completely reliable concerns the value of the ``correction'' $aR$ in
the Lagrangian $f(R)=R(1+ aR)$ and is the following. For all EOS and
for all values of the parameter $a$ ranging from $a=2. 10^{-2}$ to
$a= 10^{5}$ we found that  $|aR|$ does not exceed the value $
10^{-1}$ anywhere throughout the star. Although $|aR|<10^{-1}$ is in
the range where the perturbative approach is expected to work well,
the results that it gives, as we saw above,  can be rather different
from those obtained via the non-perturbative and self-consistent
approach.

\section{Conclusion}

In the present paper we studied models of neutron stars in
$R$-squared  gravity using a non-perturbative and self-consistent
approach. We used the fact that the $f(R)$ theories of gravity are
mathematically equivalent to a particular class of scalar-tensor
theories with nonzero potential of the scalar field. Within this
framework we constructed numerical solutions describing neutron
stars with different realistic equations of state. A wide range of
the parameter $a$ in the $R$-squared gravity was covered that is not
possible when using the perturbative approach. We had two main goals
-- to study the possible deviations from general relativity and to
compare our results with the widely used perturbative approach to
$f(R)$ gravity in order to study the possible nonlinear effects.

Our results show that the differences in the neutron star properties
induces by the $R$-squared theory of gravity can be considerable.
For example, the maximum mass of neutron stars for a specific
realistic equation of state can increase by approximately $10\%$
depending on the equation of state. Thus for example equations of
state which do not reach the two solar mass barrier in general
relativity can be reconciled with the observation if we employ
$f(R)$ theories of gravity. The neutron star radii also changes
significantly -- in general for large masses the radius increases
whereas for small masses it decreases compared to Einstein theory.
{\it But we should note that even though the deviations from general
relativity can be large, these deviations are still comparable with
the uncertainties in the nuclear matter equation of state. Therefore
the current observations of neutron star masses and radii alone can
not put constraints on the free parameter in the $R$-squared
theory}. But if the nuclear equation of state is better constrained
in the future, the current investigations would help us in
constraining $f(R)$ theories of gravity via neutron star
observations.

We addressed also in detail the comparison between the
non-perturbative  and self-consistent approach from one side and the
perurbative approach from the other. It turns out that the two
approaches lead to both qualitative and quantitative different
results, as  it can be seen from the graphs presented in the present
paper and those in \cite{Cooney09} and \cite{Arapoglu11}. The most
significant difference is the behaviour of the maximum neutron star
mass as a function of the parameter $a$ in the physical sector
$a>0$. In both the perturbative and non-perturbative approach a
decrease of the maximum mass is observed for small values of the
parameter $a$, but the quantitative differences reach large values.
In the non-perturbative approach, the decrease  is almost negligible
-- below $1\%$ for all of the realistic EOS. In contrast the results
in \cite{Arapoglu11} show a much stronger decrease of the maximum
mass. Taking into account that the maximum mass of the modern
realistic equations of state should not fall below approximately two
solar masses, constraints on the parameter $a$ were obtained in
\cite{Arapoglu11}, namely $a\lesssim 10^{6} {\rm m}^2$. It is
obviously not possible to derive such constraints in the
non-perturbative approach were the decrease of the mass is very
small and the deviations from general relativity are comparable to
the deviations coming from the use of different modern realistic
equations of state. Also in the non-perturbative approach a
considerable increase (up to approximately $10\%$) of the maximum
mass is observed for large values of $a$ and for all of the
considered EOS.

The behavior of the neutron star radius is also different -- in the
non-perturbative approach the models with larger masses have larger
radii compared to general relativity and the models with smaller
masses have smaller radii. This behaviour is exactly  opposite to
the perturbative models presented in \cite{Arapoglu11}. In addition,
new and potentially stable branches of solutions were not found
after the maximum mass for neither of the considered EOS nor values
of $a$.

Concerning the local internal structure of the stars in $f(R)$
gravity, the local Jordan frame mass increases  with the radial
coordinate for all of the considered equations of state, including
SLy4, in contrast to the observed decrease of the mass  with the
radial coordinate found using a perturbative approach in
\cite{OGPR2013}.

Concluding, the perturbative approach does not seem able to provide
reliable results for the study of the strong field regime of $f(R)$
theories and for confronting it with the observations.

In this situation, it seems that the tightest constraint on the
parameter $a$ of the $R$-squared gravity is imposed by the results
from Gravity Probe B experiment, namely  $a\lesssim 5\times 10^{11}
m^2$ \cite{Jetzer2010}. In terms of the scalar field mass $m_{\Phi}$ the constraint corresponds
to $m_{\phi} \gtrsim 10^{-13} eV/c^2$ where $c$ is the speed of light. However, this constraint is based on the weak
field regime and it  can change for more general $f(R)$ theories.

A natural generalization of the present work is to extend the
non-perturbative and self-consistent approach to  more general
$f(R)$ theories. This is numerically considerably more challenging
task and we hope to address the problem in the near future. Some of
our preliminary studies show that the nonperturbative approach gives
in the general case results different from the perturbative one.

\acknowledgments{D. D. would like to thank the Alexander von
Humboldt Foundation for support.  K.K. and S.Y. would like to thank
the Research Group Linkage Programme of the Alexander von Humboldt
Foundation for the support and S.Y. would like to thank the
Institute for Theoretical Astrophysics Tuebingen for its kind
hospitality. The support by the Bulgarian National Science Fund
under Grant No. DMU-03/6, by the German Science Foundation (DFG)
via SFB/TR7, and the networking support by the COST
Action MP1304 is gratefully acknowledged. }

\end{document}